\documentclass[conference]{IEEEtran}
\usepackage{amsmath,amssymb,amsfonts}
\usepackage{algorithmic}
\usepackage{graphicx}
\usepackage{textcomp}
\usepackage{xcolor}
\usepackage{xspace}
\usepackage{tabularx}
\usepackage{hyperref}
\usepackage{balance}
\usepackage{xcolor}
\usepackage{colortbl}

\newcommand{\etal}{et al.\xspace}
\newcommand{\ie}{i.e.\xspace}
\newcommand{\eg}{e.g.\xspace}
\newcommand{\fig}[1]{Fig.~\ref{#1}}
\newcommand{\tab}[1]{Table~\ref{#1}}

\newcommand{\figsize}{0.95\columnwidth}

\usepackage[disable]{todonotes}

\def\BibTeX{{\rm B\kern-.05em{\sc i\kern-.025em b}\kern-.08em
    T\kern-.1667em\lower.7ex\hbox{E}\kern-.125emX}}

\begin{document}
\title{On Package Freshness in Linux Distributions\\
	\thanks{Fonds de la Recherche Scientifique -- FNRS} 
}

\author{\IEEEauthorblockN{Damien Legay}
	\IEEEauthorblockA{\textit{Software Engineering Department} \\
		\textit{University of Mons}\\
		Mons, Belgium \\
		damien.legay@umons.ac.be}
	\and
	\IEEEauthorblockN{Alexandre Decan}
	\IEEEauthorblockA{\textit{Software Engineering Department} \\
		\textit{University of Mons}\\
		Mons, Belgium \\
		alexandre.decan@umons.ac.be}
	\and
	\IEEEauthorblockN{Tom Mens}
	\IEEEauthorblockA{\textit{Software Engineering Department} \\
		\textit{University of Mons}\\
		Mons, Belgium \\
		tom.mens@umons.ac.be}
}

\pagestyle{plain}
\maketitle

\begin{abstract}
The open-source Linux operating system is available through a wide variety of distributions, each containing a collection of installable software packages. It can be important to keep these packages as fresh as possible to benefit from new features, bug fixes and security patches. However, not all distributions place the same emphasis on package freshness.
We conducted a survey  in the first half of 2020 with 170 Linux users to gauge their perception of package freshness in the distributions they use, the value they place on package freshness and the reasons why they do so, and the methods they use to update packages.
The results of this survey reveal that, for the aforementioned reasons, keeping packages up to date is an important concern to Linux users and that they install and update packages through their distribution's official repositories whenever possible, but often resort to third-party repositories and package managers for proprietary software and programming language libraries.
Some distributions are perceived to be much quicker in deploying package updates than others. 
These results are valuable to assess the requirements and expectations of Linux users in terms of package freshness.
\end{abstract}

\section{Introduction}
The Linux operating system is arguably the most successful open-source project to ever have come to fruition. Per the nature of open-source software, Linux is available in many forms, called \textit{distributions}. Each distribution is composed of the Linux kernel and a host of software products provided in the form of \textit{packages}. The number of these packages in Linux distributions is known to grow superlinearly~\cite{tu2000evolution,robles2005evolution}. Packages are made available to users through a wide variety of \textit{package managers}. Some, such as \textsf{pacman}, \textsf{dkpg} or \textsf{RPM}, are specific to a distribution and its derivatives whilst others, such as \textsf{Flatpak} or \textsf{Snappy}, are intended for cross-distribution usage.
The set of packages that can be found within these package managers largely depends on the ambitions and philosophy of the distribution. Some distributions, such as \textsf{Debian}, pledge that all their components will be entirely composed of free software\footnote{https://www.debian.org/social\_contract} while other distributions make no such promise.
Philosophical divergences also cause versions of packages available in each distribution to differ, as distributions weigh concerns of \textit{stability} (the ability of a distribution to withstand changes in its components) and \textit{package freshness} (how up to date a package is compared to its upstream releases) differently.

This presents a trade-off to the maintainers of Linux distributions.
On the one hand, adopting new versions of packages within the distribution will grant users access to new features, bug fixes and security patches.
On the other hand, these new versions risk introducing breaking changes, new bugs, security vulnerabilities, incompatibilities or co-installability issues wherein packages cannot be installed without creating conflicts with some other packages~\cite{Vouillon2011, Vouillon2013}. Specifically for \textsf{Debian}, Claes~\etal estimated that the number of packages being incompatible with at least one other package oscillates between 15\% and 25\% over time~\cite{ClaesEtAl2015Debian}.
Distribution maintainers thus have to go through the very time-consuming process of assessing new versions of packages for these risks.

Reliance on semantic versioning\footnote{https://semver.org} could mitigate the cost of assessing for breaking changes but it is known that different programming language ecosystems comply to semantic versioning to different degrees~\cite{Decan2019TSE}. This implies that, depending on the language a package is written in, even patch or minor releases are likely to contain breaking changes. Specifically for Maven, Raemakers \etal~\cite{Raemaekers2017semantic} observed that about one third of all updates, including minor releases and patches, introduce API breaking changes. Similarly, the task of assessing updates for incompatibilities depends on the accuracy of the package metadata, but that metadata is often invalidated through package evolution~\cite{ArthoSCTZ12}.
These factors help explain the choice of some distribution maintainers to emphasise stability and security over package freshness.
Yet, package freshness is important to the Linux community, as evidenced by the existence of package freshness monitoring services such as \textsf{Repology} and \textsf{DistroWatch}.

Our goal is to measure, understand and compare package freshness in Linux distributions and evaluate to which extent the perception of users matches reality. To achieve this goal, we will conduct a mixed study, consisting of a qualitative component (a survey of Linux users) and a quantitative component (empirical analyses on the freshness of packages in Linux distributions). This paper constitutes a first step towards this goal, reporting upon the results of the survey. 

\section{Related Work}

Little is known about package freshness in Linux distributions in general. Shawcroft~\cite{shawcroft2009open} compared the package freshness of a limited set of 37 packages in 8 Linux distributions by looking the number and proportion of obsolete packages, measuring the time between upstream release of a package version and downstream deployment into a distribution, as well as quantifying the number of upstream versions that are ahead of deployed versions. We will extend this work to a larger corpus of packages, using more recent data.
Gonzalez-Barahona \etal~\cite{gonzalez2009macro} observed that one out of eight packages (12\%) was not updated at all
within a nine-year timespan, from \textsf{Debian Stable} 2.0 (released on 1998-07-24) to 4.0 (released on 2007-04-08). 
Nguyen and Holt~\cite{nguyen2012life} studied the lifecycle of \textsf{Debian} packages. They compared the age of packages in \textsf{Debian} distributions \textsf{Unstable}, \textsf{Testing} and \textsf{Stable}, defining package age as the time delta between its introduction into the distribution and its removal from  \textsf{Debian} or its replacement (update) by a newer version of the same package.

The freshness of distributed packages has been formalised by the notion of technical lag (expressed both in terms of time and number of versions) by Gonzalez-Barahona \etal~\cite{gonzalez2017technical,zerouali2019formal}.
Zerouali \etal~\cite{zerouali2019saner} used technical lag to explore how outdated packages are in \textsf{Debian}-based Docker containers and the impact that such outdated packages have on the presence of security vulnerabilities and bugs.
Decan \etal~\cite{decan2018evolution} used it to assess the reluctance of package maintainers to update package dependencies in order to avoid putative backward-incompatible changes in programming language ecosystems.

\section{Methodology}
This paper reports on a survey of Linux users, conducted in early 2020, aiming to examine their perception of package freshness. We specifically explore the value Linux users place on package freshness, their motivations to upgrade packages to newer versions, the way they use to do so and how fresh they perceive the packages in their most used distribution to be.
In order to obtain a sampling of the views of the open-source community at large, we distributed the survey to attendants of the Free and Open source Software Developers' European Meeting (FOSDEM 2020) and Community Health Analytics Open Source Software conference (CHAOSSCon Europe 2020), in both paper and electronic versions.
We received 68 responses, 52 from CHAOSSCon and FOSDEM, 9 from convenience sampling and 7 from computer science students at our university. 37 answers came in paper form and 31 were submitted online.
We also posted the survey on Twitter and Linux-related fora and subreddits, obtaining an additional 102 responses, for a total of 170. The fora were: \textsf{linux.org}, \textsf{forums.fedoraforum.org}, \textsf{forums.debian.net}, \textsf{forums.linuxmint.org} and the subreddits \textsf{/r/centos}, \textsf{/r/fedora}, \textsf{/r/redhat}, \textsf{/r/linuxmint} and \textsf{/r/opensuse}. We did not receive authorisation to post the survey on other Linux-related fora and subreddits.
The survey form and anonymised answers can be found at \url{https://doi.org/10.5281/zenodo.3908332}.

Since the answers to the survey questions can depend on the distribution(s) used, we asked respondents which Linux distribution(s) they frequently make personal use of. They could rank up to three distributions, in order of frequency of use, chosen from a pre-established list of 16 popular Linux distributions. They also had the option to specify other distributions.

\tab{distroRankingsTable} reports the total number of answers obtained for each distribution, as well as the number of times that distribution was ranked first, second or third. The table also reports the aggregated responses for each family of Linux distributions. Distributions for which there were fewer than five answers have been gathered under the label ``other distributions''. These are \textsf{Gentoo}, \textsf{SUSE Entreprise Edition}, \textsf{Parabola}, \textsf{FerenOS}, \textsf{Android}, \textsf{Clear Linux}, \textsf{Knoppix}, \textsf{Alpine Linux}, \textsf{NixOS} and \textsf{Raspbian}.
These results indicate that 88\% (149) of the respondents make use of at least two distributions, and 62\% (106) of at least three, showing that most of them have experience with several distributions, either from using them concurrently or migrating from one to another.
Often, people who use more than one distribution use one that favours stability (\eg \textsf{CentOS}) and one that favours freshness (\eg \textsf{Fedora}).

\begin{table}[!tb]
	\caption{Usage per (family of) Linux distribution(s).}
	\label{distroRankingsTable}
	\centering
	\begin{tabular}{l|c|c|c|c}
		\textbf{Distribution} & \textbf{First} & \textbf{Second} & \textbf{Third} & \textbf{Total} \\
		\hline
		\hline
		\rowcolor{gray!25}
		\textbf{Ubuntu} (family) & 47 & 46 & 43 & 117 \\
		\quad Ubuntu LTS & 30 & 25 & 11 & 66 \\
		\quad Ubuntu & 17 & 16 & 18 & 51 \\
		\rowcolor{gray!25}
		\textbf{Debian} (family) & 30 & 37 & 26 & 93 \\
		\quad  Stable & 19 & 24 & 22 & 65 \\
		\quad  Testing & 10 & 13 & 4 & 27 \\
		\quad  Unstable & 1 & 0 & 0 & 1 \\
		\rowcolor{gray!25}
		\textbf{Red Hat} (family) & 33 & 33 & 25 & 91 \\
		\quad Fedora & 24 & 9 & 8 & 41 \\
		\quad CentOS & 8 & 16 & 9 & 33 \\
		\quad Entreprise Edition & 1 & 8 & 8 & 17 \\
		\rowcolor{gray!25}
		\textbf{Arch Linux} & 29 & 8 & 8 & 45 \\
		\rowcolor{gray!25}
		\textbf{OpenSUSE} (family) & 21 & 13 & 5 & 39 \\
		\quad Tumbleweed & 17 & 4 & 2 & 23 \\
		\quad LEAP & 4 & 9 & 3 & 16 \\
		\rowcolor{gray!25}
		\textbf{Linux Mint} & 5 & 10 & 7 & 22 \\
		\rowcolor{gray!25}
		\textbf{Slackware} & 2 & 2 & 1 & 5 \\
		\rowcolor{gray!25}
		\textbf{Other distributions} & 3 & 6 & 5 & 14 \\
		\hline
		\textbf{Total} & 170 & 149 & 106 & \\
	\end{tabular}
\end{table}
\section{Findings}
\subsection{Which distributions are perceived to be more up-to-date?}
\label{perceptionSection}
To gauge the user perception of package freshness in Linux distributions, we asked respondents how long it took, according to them, for the latest upstream version to be made available in the official repositories of their most-used distribution (answered first in the previous question). We gave them six exclusive options: ``a few days, at most'', ``a few weeks'', ``a few months'', ``a few years'', ``never (not available)'' and ``I don't know''.
We asked about six categories of packages:\\
\emph{OSS: open-source end-user software} (\eg \textsf{Firefox} or \textsf{GIMP});\\
\emph{PS: proprietary end-user software} (\eg \textsf{Adobe Reader}, \textsf{Spotify} or \textsf{Skype});\\
\emph{DT: development tools} (\eg \textsf{git}, \textsf{Emacs} or \textsf{Eclipse});\\
\emph{STL: system tools and libraries} (\eg \textsf{openSSL}, \textsf{sudo} or \textsf{zsh});\\
\emph{PLL: programming language libraries} (\eg \textsf{NumPy} for Python, \textsf{Lodash} for npm)\\
\emph{PLR: programming language runtimes} (\eg \textsf{Python}, \textsf{node.js} or \textsf{Java}).

\begin{figure}[!tb]
	\centering
	\begin{tabular}{c}
		\includegraphics[width=\figsize]{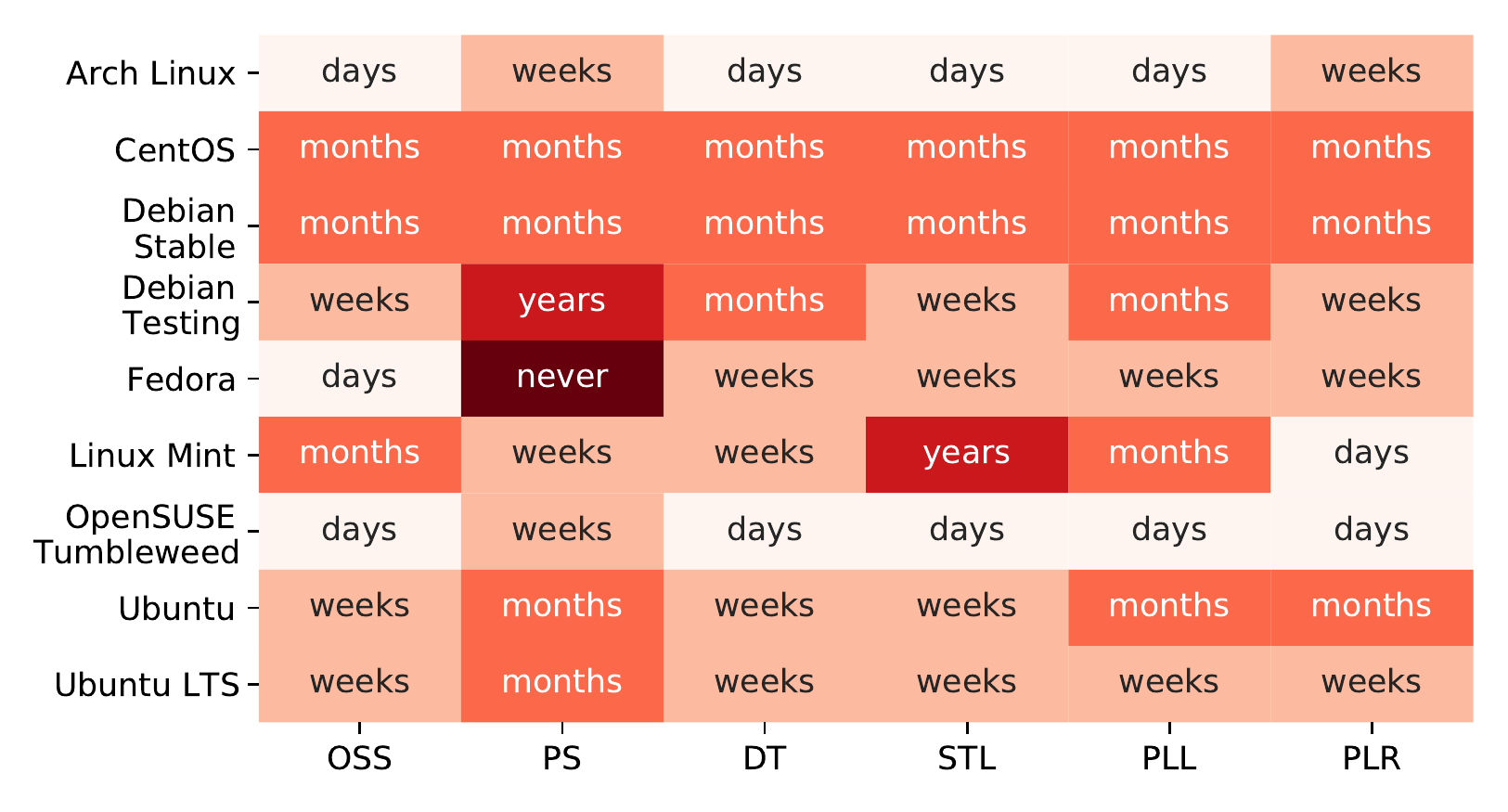}
	\end{tabular}
	\caption{Package freshness perception.}
	\label{perceptionMedians}
\end{figure}

\fig{perceptionMedians} presents the median answer for each category. Only distributions which are used primarily by five or more respondents are shown.

Being based on a rolling release policy, \textsf{Arch Linux} and \textsf{OpenSUSE Tumbleweed} strive to distribute the latest stable releases of all packages included in the distribution. \fig{perceptionMedians} reveals that the respondents' perception aligns with this reality, since most of them agree that upstream versions are made available very quickly, within days. Only proprietary end-user software (PS) packages are perceived by some to be updated slower than a matter of weeks, when at all available, in accordance with the fact that some distributions do not directly support proprietary software.

\textsf{Fedora} users believe that they will dispose of fresh versions within weeks. Most \textsf{Fedora} users answered that proprietary software just was not available in Fedora's official repositories. \textsf{Ubuntu}, \textsf{Linux Mint} and \textsf{Debian Testing} users usually think it takes weeks to months for upstream versions to be released within the official repositories. Although the median answer for proprietary software in \textsf{Debian Testing} is that it takes years, this is due to the fact that a significant portion of respondents answered proprietary software was not available in \textsf{Debian Testing}. At the end of the spectrum, \textsf{CentOS} and \textsf{Debian Stable} users tend to expect to wait months for fresh versions to be made available in their distribution's official repositories, regardless of package type.
7 out of the 170 respondents expressed that they did not know when updates are made available to the distribution's official repositories.
A further 50 respondents expressed ignorance for only some categories, principally regarding proprietary software.

\subsection{To what extent do users value package freshness?}
\label{valueSection}

\begin{figure}[!tb]
	\centering
	\begin{tabular}{c}
		\includegraphics[width=\figsize]{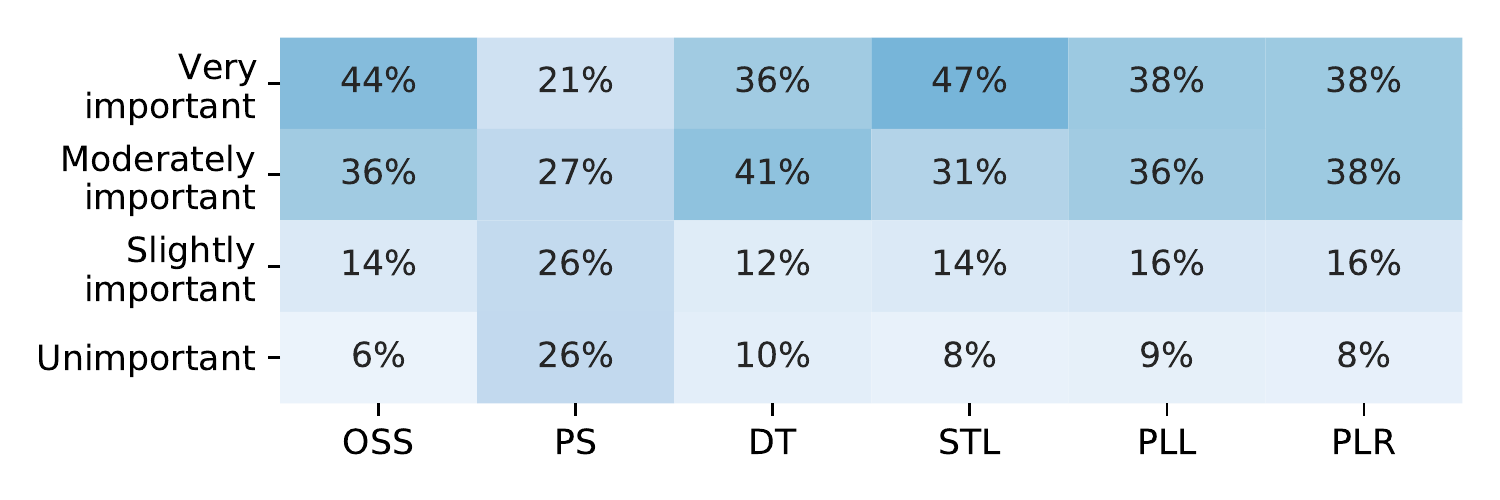}
	\end{tabular}
	\caption{Importance imparted by respondents to staying up-to-date}
	\label{importance}
\end{figure}

We enquired, for each package category, what importance users impart to keeping packages up to date with upstream releases.
To do so, we relied on a 4-value Likert scale to denote relative importance: unimportant, slightly important, moderately important and very important.
\fig{importance} reports on the results for each package category.

For all categories, respondents consider it important to update packages: 75\% to 80\% of them answered it was moderately to very important to remain up to date.
A notable exception is the proprietary end-user software category (PS): in this instance alone, a majority (52\%) considers the importance of package freshness to be slight or null.
We do not see a clear practical reason why users would consider updating proprietary packages less important than other packages.

Users of distributions that are perceived to be slower in deploying packages, such as \textsf{CentOS} and \textsf{Debian Stable}, were less likely to value maintaining a high level of package freshness. Conversely, users of distributions that are perceived to have fresher packages, such as \textsf{Arch Linux}, \textsf{OpenSUSE Tumbleweed} and \textsf{Fedora} were more likely to consider package freshness important.
Indeed, across all categories, a much greater proportion of respondents answered that updating packages was moderately to very important for \textsf{Arch} (80\%), \textsf{Tumbleweed} (80\%) and \textsf{Fedora} (84\%), than for \textsf{CentOS} (50\%) and \textsf{Debian Stable} (54\%).

\subsection{What are the main reasons for updating packages?}
Benefits to updating packages include access to new features, bug fixes and security patches.
In order to verify whether those benefits actually motivate users to update, respondents were asked what their main reasons were to update packages, out of five options: to benefit from security patches (selected by 90\% of the respondents), from bug fixes  (80\%), from new features (66\%), to sate their desire to remain up to date (35\%) or to retain compatible with other packages (27\%).
They could select as many options as they wanted. An additional open option was available, but not used by any respondent. The motivation of obtaining new features is less prevalent in users of distributions that are perceived as less fresh, such as \textsf{Debian Stable} and \textsf{CentOS}. Users of \textsf{Debian} are less likely to cite bugs as reasons to update, likely owing to the Debian process of package integration leading to stable distributions, particularly \textsf{Debian Stable}.

\subsection{Which mechanisms are used to keep packages up-to-date?}
Several mechanisms can be used to update packages: using the \emph{official package manager of the distribution and its official repository (\textsf{off})}, using the \emph{official package manager of the distribution with community repositories (\textsf{com})}, using \emph{third-party package managers such as \textsf{Flatpak} (\textsf{3rd})}, \emph{installing manually from binaries (\textsf{bin})} or \emph{installing manually from source files (\textsf{src})}.

\begin{figure}[!tb]
	\centering
	\begin{tabular}{c}
		\includegraphics[width=\figsize]{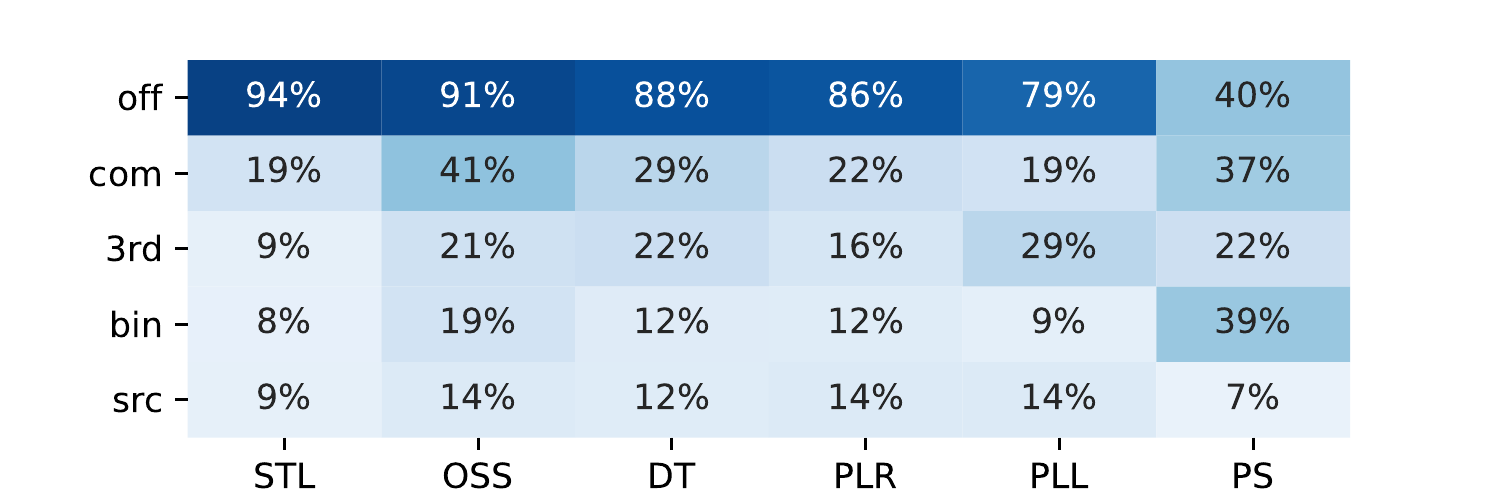}
	\end{tabular}
	\caption{Frequency of updating mechanism usage.}
	\label{mechanisms}
\end{figure}

We asked respondents which of these mechanisms they used to update packages in their most-used distribution, for each of the considered package categories. They could select as many answers as they wanted.
\fig{mechanisms} shows a heatmap of the frequency at which respondents reported which of these mechanisms is used, for each package category.

For most package categories, the official repositories (\textsf{off}) dominate largely (used by 79\% to 94\% of respondents), followed by community repositories (used by 19\% to 41\%).
Proprietary software stands in contrast to other categories, being installed more than a third of the time from binaries (\textsf{bin}). This is expected, as some distributions (\eg \textsf{Debian}) are reluctant to include proprietary software within their official repositories.
We also see that, despite their recency, third-party managers (\textsf{3rd}) such as \textsf{Flatpak} or \textsf{Snappy} are regularly used to install end-user open source software (21\%), development tools (22\%) or programming language libraries (29\%).
We also observe that programming language libraries are less often installed through official repositories, and are installed nearly one third of the times through specific third-party package managers. This should be no surprise given that most libraries for these languages are (sometimes exclusively) available through dedicated package managers (\eg \textsf{pip} for \textsf{Python}, \textsf{npm} for \textsf{Javascript}).
In the case of packages related to development tools, we believe this is likely due to the assortment of tools available through \textsf{Flatpak} and \textsf{Snappy}, including popular IDEs, text editors and graphical user interfaces for \textsf{git}.
These tools are not always available in official or community repositories.
For instance, \textsf{Intellij IDEA} is not available in the official repositories of Fedora, but can be installed through \textsf{Flatpak} and \textsf{Snappy}.
\section{Discussion}
We reported the results of a survey of 170 Linux users about package freshness.
We found that users perceive distributions such as \textsf{Arch Linux}, \textsf{OpenSUSE Tumbleweed} and \textsf{Fedora} as being much more likely to have fresh packages than distributions such as \textsf{Debian Stable} and \textsf{CentOS}. Verifying these perceptions will require a quantitative empirical comparison of freshness in Linux distributions.

\begin{figure}[!tb]
	\centering
	\begin{tabular}{c}
		\includegraphics[width=\figsize]{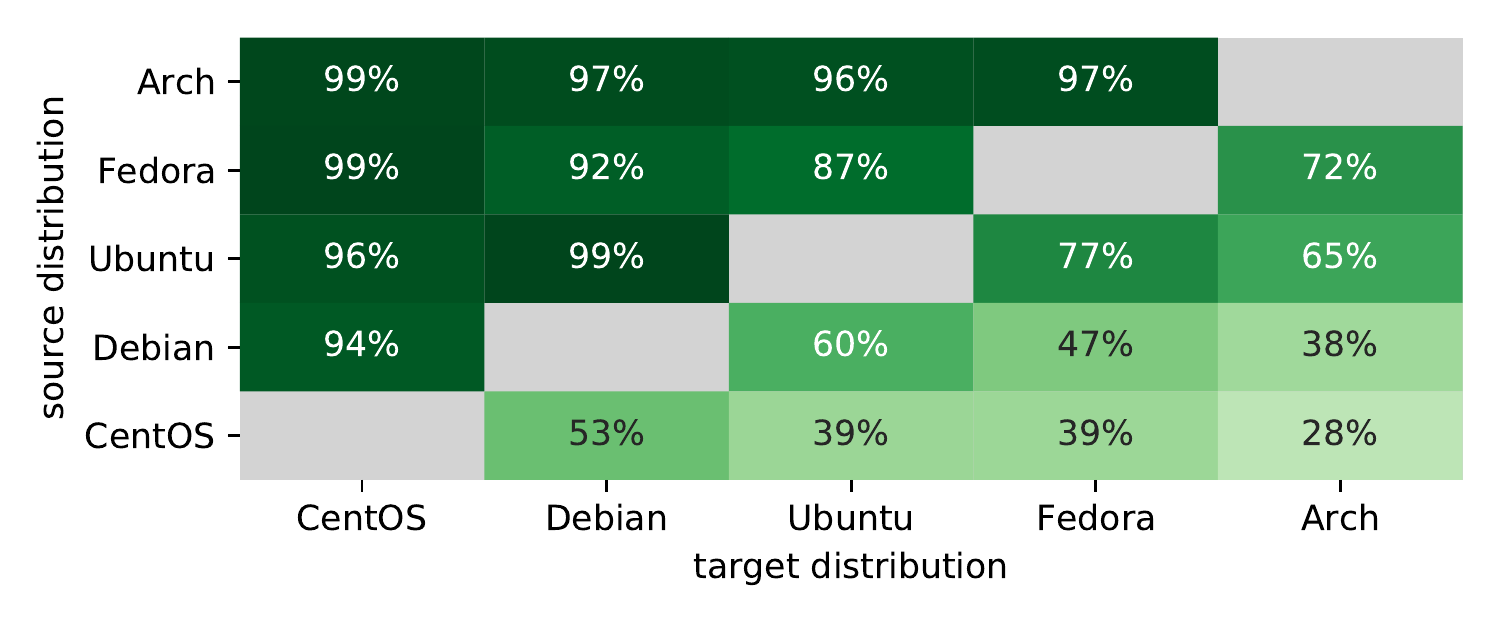}
	\end{tabular}
	\caption{Proportion of packages in a source distribution that are at least as fresh as in a target distribution}
	\label{fresherHeatmap}
\end{figure}

As a preliminary step towards such an empirical study, we gathered the package versions available in a recent snapshot of the five distributions respondents cited the most. We took the latest snapshot available prior 2019-11-01. This date was selected to minimise the gaps between distribution release dates. \textsf{Ubuntu 19.10} was chosen over \textsf{LTS} versions of the distribution for that reason. The distributions considered are listed in \tab{outdatedTable}. We identified 529 common packages for these distributions. We performed pairwise comparisons on the freshness of the versions of these packages present in those distributions. The results are found on \fig{fresherHeatmap} with each cell reporting the proportion of packages that are at least as fresh in the source distribution as in the target distribution. For instance, 99\% of packages in \textsf{Arch Linux} are at least as fresh as those in \textsf{CentOS}, whereas only 28\% of packages in \textsf{CentOS} are at least as fresh as those in \textsf{Arch Linux}. This means that 72\% of the packages in \textsf{CentOS} are outdated with respect to those available in \textsf{Arch Linux}.

\begin{table}[!tb]
	\caption{Releases of the considered Linux distributions}
	\label{outdatedTable}
	\centering{
	\begin{tabular}{l|c|c}
		\textbf{Distribution} & \textbf{Release} & \textbf{Date} \\
		\hline
		\hline
		Arch Linux & rolling & 2019-10-31 \\
		CentOS & 8.0 & 2019-09-24 \\
		Debian Stable & 10 & 2019-07-06 \\
		Fedora & 31 & 2019-10-29 \\
		Ubuntu & 19.10 & 2019-10-17 \\
	\end{tabular}
	}
\end{table}

We see that almost all packages in \textsf{Arch Linux} are at least as fresh as the packages found in the other distributions, and only 28\% (\textsf{CentOS}) to 72\% (\textsf{Fedora}) of packages in other distributions are as fresh as \textsf{Arch Linux} packages. At the other extreme, the vast majority of packages (94\% or more) in all considered distributions are at least as fresh as those found in \textsf{CentOS} and only 28\% to 53\% of \textsf{CentOS} packages are at least as fresh as those found in other distributions. \textsf{Debian Stable} is in a similar situation as \textsf{CentOS}, albeit slightly less marked.
\textsf{Fedora} and \textsf{Ubuntu} lie in the middle, with 72\% and 65\% (resp.) of their packages as fresh as those found in \textsf{Arch}.

These results suggest the following ranking of distributions in decreasing order of freshness: \textsf{Arch Linux}, \textsf{Fedora}, \textsf{Ubuntu}, \textsf{Debian Stable} and finally \textsf{CentOS}. This roughly corresponds to respondent perception in \fig{perceptionMedians}.
Nevertheless, these perceptions are imprecise, as evidenced by the fact that \textsf{Ubuntu LTS} users considered some categories of packages to be available within their distribution's repositories sooner than non-LTS \textsf{Ubuntu} users, which is contrary to expectations.
Additionally, 57 respondents (33\%) were not confident to answer for at least one package category, with 7 (4\%) of them answering they did not know for all categories.

This preliminary empirical analysis of package freshness in five Linux distributions hints to the fact that distributions lie on a continuum with regards to the trade-off between package freshness and system stability.
In a follow-up study, we will seek to empirically quantify the difference in package freshness between distributions by relying on the technical lag measurement framework~\cite{zerouali2019formal}. We will compare the versions of packages available in different distributions in terms of time lag (\ie the time since a more recent upstream version of the package has been available) and version lag (\ie the number of more recent versions available).
This will allow us to place distributions on that continuum, helping users choose a distribution that best matches their expectations in terms of freshness and stability.
We will seek to contrast the package freshness measured within distributions with the perceptions of users reported in this paper, thereby gauging to what degree user perception matches reality.

We will also examine the relationship between package freshness, stability and security in distributions.
Comparing distributions in terms of these characteristics could help users to choose a distribution that matches their requirements and expectations.
It will also allow package maintainers to know whether their packages are likely to be up to date in certain distributions, and potentially adopt practices that allow distribution maintainers to assess the stability, compatibility and security of their packages more quickly, to allow faster deployment of updates.
Additionally, we will conduct a follow-up survey to examine to what extent package freshness, stability and security motivate users to migrate from one distribution to another.

\section{Conclusion}
The Linux ecosystem depends on a set of packages. These packages are made available through package managers, either official ones used by specific distributions, or third-party ones, as well as directly through binaries and source files. Package versions available in official distribution repositories do not always match the latest versions released by the package's authors, out of a need to balance package freshness with system stability and security.

This paper is a first step towards a mixed study to understand, measure and compare package freshness in Linux distributions.
We reported on the results of a survey aimed at habitual Linux users to determine what were their values, perceptions and practices regarding package freshness. Their answers indicated that they usually place significant value in keeping the packages they use up to date, principally out of security concerns, but also largely to benefit from bug fixes and new features.
Whenever possible, users prefer to update packages through the distribution's official package managers, using the distribution's official repositories. This is not always possible, though, as some packages are either unavailable or outdated in official repositories.
This is most prevalent in the case of proprietary end-user software and some development tools. Additionally, programming language libraries are often installed and updated through language-specific package managers.
Unsurprisingly, users perceive packages in rolling release distributions to be very fresh. On the other hand, \textsf{Debian Stable} and \textsf{CentOS} users perceive it takes longer for new versions of packages to be made available in the official repositories, in the order of months. Other distributions are perceived to be somewhere in between on the ``package freshness continuum''.

Preliminary empirical analysis shows that there is some truth to this perception, with \textsf{Arch Linux} being the most fresh distribution studied and \textsf{CentOS} the least. In a follow-up work, we will conduct further empirical analyses in order to quantify the comparative package freshness of Linux distributions, as well as examine its role as a motivator in user adoption of distributions.

\section*{Acknowledgment}
This research is supported by the Fonds de la Recherche Scientifique -- FNRS under Grants number O.0157.18F-RG43 (Excellence of Science project SECO-ASSIST) and T.0017.18.
\newpage
\bibliographystyle{IEEEtran}
\balance
\bibliography{references}

\end{document}